\documentclass[pre,twocolumn]{revtex4}
\usepackage{amsmath,amsfonts}

\usepackage{graphicx}
\usepackage{color}

\usepackage{dcolumn}% Align table columns on decimal point
\usepackage{bm}% bold math

\usepackage{units}

\def\bf{\textbf}

\begin{document}

%\begin{CJK*}{GBK}{kai}

\title{Causal diffusions, causal Zeno effect and collision number}

\author{Ji-Rong Ren}%(ÈμÌÈÙ)}
\email{renjr@lzu.edu.cn}

\author{Ming-fan Li\footnote{Corresponding author.}}%(ÀîÃ÷·²)}
\email{limf07@lzu.cn}

\author{Tao Zhu}%(ÖìÌÎ)}
\email{zhut05@lzu.cn}

\affiliation{Institute of Theoretical Physics, Lanzhou University,
Lanzhou, 730000, China}

\begin{abstract}
We consider diffusion processes with the help of Markov random walk
models. Especially the process of diffusion of a relativistic
particle in a relativistic equilibrium system is considered. We
interpret one of the results as causal Zeno effect for its
similarity to quantum Zeno effect. Another problem we considered is
about collision number. Basing on our numerical results, we propose
that in the considered situation the probability density
distribution among different collision numbers is a lognormal
distribution.

\end{abstract}

\maketitle

\section{Introduction}

Diffusion process is very common in physics, chemistry, biology and
many other fields. It can be described by a diffusion equation with
or without a finite maximum velocity
\cite{CG,ExtIrrThermo,GCE,Chaos}. Another approach toward diffusion
process is numerical simulation \cite{Cubero,2D,statistic}. Yet, it
can also be considered with the help of random walk models (RWM).
Many literatures take this way, such as
\cite{DunkelRWM,PhysRep,PRW}.

In this letter, we consider the diffusion of a particle in an
equilibrium system in the framework of Markov random walk model. We
begin our study with the one-particle velocity distribution of a
non-relativistic/relativistic equilibrium system.

As it is well-known, the one-particle velocity distribution of a
near-free non-relativistic equilibrium gas is the Maxwell's
probability density function (PDF):
\begin{equation}\label{MaxwePDF}
    f_M(\textbf{v};m,\beta)=(\frac{\beta m}{2\pi})^{d/2}e^{-\beta
    m\textbf{v}^2/2}.
\end{equation}
The first relativistic generalization of Maxwell's function was
proposed by F. J\"{u}ttner from a consideration of maximum entropy
principle \cite{Juttner,NJP,ZygPLA}:
\begin{equation}\label{JuttnerP}
    f_{Jp}(\textbf{p};m,\beta)=\frac{1}{Z_J}e^{-\beta E},
\end{equation}
where $E=\sqrt{m^2+\textbf{p}^2}=m\gamma(\textbf{v})$,
$\textbf{p}=m\textbf{v}\gamma(\textbf{v})$, and
$\gamma(\textbf{v})=1/\sqrt{1-\textbf{v}^2}$. $Z_J$ is the
normalization factor. Written in velocity PDF, J\"{u}ttner's
distribution is
\begin{equation}\label{JuttnerV}
    f_{Jv}(\textbf{v};m,\beta)=\frac{m^d}{Z_J}\gamma(\textbf{v})^{2+d}e^{-\beta m\gamma(\textbf{v})}.
\end{equation}

A lot of work has been done basing on J\"{u}ttner's distribution
since its proposal. However, in the 80's of the last century, doubt
about J\"{u}ttner's function was expressed and ``modified
J\"{u}ttner's function" was proposed \cite{Horwitz1,Horwitz2,NJP}.
But recent numerical simulations favored J\"{u}ttner distribution
(\ref{JuttnerV}) as the correct one-particle velocity distribution
of a relativistic equilibrium system \cite{Cubero,2D}.

So, in this letter Maxwell's function (\ref{MaxwePDF}) and
J\"{u}ttner's function (\ref{JuttnerV}) are used. Numerical analysis
is done. By comparing the results from the two distributions, we
interpret one of the results as causal Zeno effect for its
similarity with the Quantum Zeno Effect \cite{QZE,wikiQZE}.

Another problem we consider is that how many times a particle will
collide with other particles during the process of diffusion from a
initial point to a final point in a given time interval. As one can
conceive, there will be a probability distribution among different
collision numbers. One of our results is that if the final point is
not specified, the probabilities to collide whatever N times will be
the same. For the case of that the final position is also fixed, we
make numerical analysis and propose that the probability
distribution is the lognormal distribution \cite{CLT,LND,wikiLND}.

\section{Random walk models}\label{RWM}

In this section, we consider diffusions as random walk processes. At
first we give some general descriptions, then we apply the model to
the non-relativistic Markov diffusions, finally we consider the
relativistic Markov diffusions. Our discussions will be confined to
the one dimensional case.

\subsection{General descriptions}

When a particle is diffusing in an environment, one wants to know
the probability for it to reach some position from a given position
within a given time interval. The formal kinematics formula is
\begin{equation}\label{FKF}
    \xi(t)=x_0+\int_0^t ds v(s),
\end{equation}
where $v(s)$ is its velocity.

For a free particle, its velocity is a constant. While if it
interacts with other matters, its velocity will change. Here, we
assume that the particle interacts with other particles only by
point collision. Thus the above formula transforms into
\begin{equation}\label{FKF_D}
    \xi(t)=x_0+\sum_{i=1}^N v_i(t_i-t_{i-1}),
\end{equation}
where $t_i,i=1,2,...,N-1$ are the times of collisions and $t_N$ is
the given ending time. We write $\tau_i=t_i-t_{i-1}$ for simplicity.
Although the actual collision number is not $N$ but $N-1$, one can
roughly say it is or it is indicated by $N$.

$(\{v_i\};\{\tau_i\})$ defines a path in the space. If the particle
has a probability density $f(\{v_i\};\{\tau_i\})$ to follow this
path. Then the transition probability density function of order N
can be written as
\begin{eqnarray}
  p_N(t,x|0,x_0)=\int dv_1...dv_N\int d\tau_1...d\tau_N \nonumber \\
    f(\{v_i\};\{\tau_i\})\delta(x-\xi(t)).
\end{eqnarray}

If the collision number is not concerned about, the total transition
PDF is
\begin{equation}\label{Total TPDF}
    p(t,x|0,x_0)=\sum_{N=1}^\infty p_N(t,x|0,x_0).
\end{equation}

In the following, we set $\tau_i$ be equal ($=\tau=t/N$). And we
consider Markov process in which case $f(\{v_i\};\{\tau_i\})$ can be
written as $\prod_{i=1}^N f(v_i;\tau_i)$.

Thus the N-order transition PDF is
\begin{equation}\label{NTPDF}
    p_N(t,x|0,x_0)=\bigg[\prod_{i=1}^N \int dv_i f(v_i;\tau_i)\bigg]\delta(x-\xi(t))
\end{equation}

Using
\begin{equation}\label{Delta}
    \delta(x)=\frac{1}{2\pi}\int_{-\infty}^\infty dk e^{ikx}
\end{equation}
together with (\ref{FKF_D}), one can get
\begin{eqnarray}
%  to remove numbering (before each equation)
   p_N(t,x|0,x_0) = \frac{1}{2\pi}\int_{-\infty}^\infty dk e^{ik(x-x_0)}\bigg[\prod_{i=1}^N \int dv_i f(v_i;\tau)e^{-ik\tau
   v_i}\bigg] \nonumber
\end{eqnarray}
\begin{equation}\label{pNfv}
% \nonumber to remove numbering (before each equation)
   = \frac{1}{2\pi}\int_{-\infty}^\infty dk e^{ik(x-x_0)}\varphi(-k\tau)^N
\end{equation}
where $\varphi(u)$ is the characteristic function of $f(v)$, i.e.
its Fourier transformation.

One can also get an expression for $p_N(t,x|0,x_0)$ by invoking the
mean velocity. Define
\begin{equation}\label{V_SN}
    V_{SN}=\sum_1^N v_i,
\end{equation}
then the mean velocity
\begin{equation}\label{V_MN}
    \overline{V}_N=\frac{1}{t}\sum_1^N v_i\tau_i=\frac{1}{N}\sum_1^N
    v_i=\frac{1}{N}V_{SN}.
\end{equation}

If the velocity probability density function of $v_i$ is $f_i(v)$,
then the PDF of $V_{SN}$ is
\begin{equation}\label{PDF_V_SN}
    f_{SN}(V)=f_1(v)*f_2(v)*...*f_N(v),
\end{equation}
here $*$ means convolution.

This leads to the probability density function of $\overline{V}_N$:
\begin{equation}\label{PDF_V_MN}
    f_{MN}(V)=f_{SN}(NV)N.
\end{equation}

Finally the N-order transition PDF is
\begin{equation}\label{NTPDF_V_MN}
    p_N(t,x|0,x_0)=f_{MN}(\frac{x-x_0}{t})\frac{1}{t}.
\end{equation}

We will follow this way when we make numerical analysis.

\subsection{Non-relativistic Markov diffusions}

Firstly we apply this model to the case of non-relativistic Markov
diffusions. The equilibrium velocity PDF of a non-relativistic
system is the Maxwell distribution.

Substituting (\ref{MaxwePDF}) into (\ref{pNfv}), and assuming
\begin{equation}\label{beta_tau_maxwell}
    \beta=\frac{1}{k_BT}=\frac{\tau}{2mD},
\end{equation}
one gets
\begin{eqnarray}\label{p_N_maxwell}
% \nonumber to remove numbering (before each equation)
  p_N(t,x|0,x_0) &=& \frac{1}{\sqrt{4\pi DN\tau}}e^{-\frac{(x-x_0)^2}{4DN\tau}} \nonumber \\
   &=& \frac{1}{\sqrt{4\pi Dt}}e^{-\frac{(x-x_0)^2}{4Dt}},
\end{eqnarray}
which is the solution to the ordinary diffusion equation
\cite{CG,ExtIrrThermo,GCE,Chaos} with a proper initial condition.

One can easily see that this expression of transition PDF is in
conflict with the special relativity, for when $|x-x_0|>ct$, there
is still a small but non-vanishing probability for the particle to
be found.

Another property of non-relativistic Markov diffusions is that for
different values of N, the $p_N(t,x|0,x_0)$'s are the same, i.e. no
matter how much the collision number is, even when it is infinite,
$p_N(t,x|0,x_0)$ is the same as above. So, the particle can always
diffuse. As we will see, this is not the case in relativistic Markov
diffusions.

\subsection{Relativistic Markov diffusions}

\subsubsection{Formula development}

For relativistic Markov diffusions, one should use J\"{u}ttner's
distribution (\ref{JuttnerV}) instead of Maxwell's distribution.
Then the N-order transition PDF will be
\begin{eqnarray}
% \nonumber to remove numbering (before each equation)
  &&p_N(t,x|0,x_0) \nonumber \\
  =&& \frac{1}{2\pi}\int_{-\infty}^\infty dke^{ik(x-x_0)}\bigg[\int dv f_J(v;\tau)e^{-ik\tau
  v}\bigg]^N \nonumber\\
  =&& \frac{1}{2\pi}\int_{-\infty}^\infty dk e^{ik(x-x_0)}\varphi_J(-k\tau)^N.
\end{eqnarray}

This is very difficult, or perhaps, even impossible to carry out the
integral analytically. For a compensation, we consider its large-N
asymptotic behavior by invoking its N-order mean velocity
$\overline{V}_N$.

The N-order mean velocity during the time interval from 0 to t is
\begin{equation}\label{MeanV_N}
    \overline{V}_N=\frac{\xi(t)-x_0}{t}=\frac{1}{t}\sum_{i=1}^N \tau_i
    v_i=\frac{1}{N}\sum_{i=1}^N v_i.
\end{equation}

Since $M(v_i)=0$, $Var(v_i)=\sigma^2<\infty$, Central Limit Theorem
(CLT) \cite{CLT} asserts that the distribution of $\sum_{i=1}^N
v_i/\sigma\sqrt{N}$ converges to Gauss's normal distribution with
parameters (0,1):
\begin{equation}
    N(x;0,1)=\frac{1}{\sqrt{2\pi}}e^{-\frac{x^2}{2}}.
\end{equation}

Then, the probability density about $\overline{V}_N$ will be
\begin{equation}
    \frac{1}{\sqrt{2\pi\sigma^2/N}}e^{-\frac{\overline{V}_N^2}{2\sigma^2/N}}.
\end{equation}

And the N-order transition PDF will be
\begin{equation}\label{NTPDF_CLT}
    p_N(t,x|0,x_0)=\frac{1}{\sqrt{2\pi t\tau\sigma^2}}e^{-\frac{(x-x_0)^2}{2t\tau\sigma^2}}
\end{equation}

Here $\sigma^2$ is the variance of J\"{u}ttner' distribution
(\ref{f_J_1}); it is a function of $\chi(=m\beta)$, and $\tau=t/N$.

At a given reciprocal temperature $\beta$, $\sigma^2$ is fixed. When
$N\rightharpoonup \infty$, $\tau \rightharpoonup 0$. Then
\begin{equation}\label{p_N_Delta}
    p_N(t,x|0,x_0)\rightharpoonup \delta(x-x_0).
\end{equation}

This result means that the particle will not diffuse at all in this
situation. By comparing the results in non-relativistic and
relativistic situations, one can see that they are contrary to each
other. In the non-relativistic situation, we have seen that
$p_N(t,x|0,x_0)$ is independent of N, and always is a Gaussian
distribution. But here, in the relativistic case, we have the result
(\ref{p_N_Delta}).

\subsubsection{Causal Zeno effect}

We venture to interpret the result (\ref{p_N_Delta}) as causal Zeno
effect for its similarity with the quantum Zeno effect. Quantum Zeno
effect is a name coined by George Sudarshan and Baidyanath Mishra in
1977 in their analysis of the situation in which an unstable
particle, if observed continuously, will never decay \cite{QZE}. One
can nearly ``freeze" the evolution of the system by measuring it
frequently enough in its (known) initial state \cite{wikiQZE}.

In the random walk model, $N\rightharpoonup\infty$ means that the
particle is continuously collided by other particles. In other
words, it is continuously observed. So the result (\ref{p_N_Delta})
means that the particle will never depart from its initial position
if it is continuously observed. However, as we have seen, in the
non-relativistic situation, the particle can always diffuse.

This difference between the two situations arises from the different
velocity distribution functions. Furthermore, as one can see, when
$\chi$ becomes large, J\"{u}ttner's distribution (\ref{JuttnerV})
tends to Maxwell's distribution, save for its finite support. Thus
the boundedness of velocity is indispensable for the convergence to
a delta function.

One can also perceive that at a given temperature, there will be a
probability distribution among different values of N, and the
behavior of $p_N(t,x|0,x_0)$ with the most possible value of N will
dominate in the diffusion process of the particle.

\subsubsection{$\lambda_N=1$}

As for the probability distribution among different values of N, we
first consider the following expression:
\begin{equation}\label{lamdaN}
    \lambda_N\equiv\int_{-\infty}^\infty p_N(t,x|0,x_0)dx.
\end{equation}
This expression quantifies the probability of the particle colliding
$N-1$ times with other particles in a given time interval $t-0$ from
the initial position $x_0$ to non-specified final positions.

Substituting (\ref{pNfv}) into the above expression, the outcome is
\begin{equation}
  \lambda_N = \int_{-\infty}^\infty dx \frac{1}{2\pi}\int_{-\infty}^\infty dk e^{ik(x-x_0)}\varphi_J(-k\tau)^N.
\end{equation}

As
\begin{equation}
    \delta(k)=\frac{1}{2\pi}\int_{-\infty}^\infty dxe^{ik(x-x_0)},   \nonumber
\end{equation}
one gets
\begin{equation}
    \lambda_N=\varphi_J(0)^N=1, \nonumber
\end{equation}
which means that the probabilities for the particle to collide
different $N-1$ times with other particles are the same. And it is
right even for cases that $\tau_i$ (i=1,2,...,N) are different and
the process is non-Markovian.

It is surprising as we have thought that there would be a
non-trivial probability distribution. However, in the case with a
specified final position, one may anticipate the probabilities will
be different. We will check this by numerical analysis.

\section{Numerical analysis}

\begin{figure}
  % Requires \usepackage{graphicx}
  \includegraphics[width=7cm]{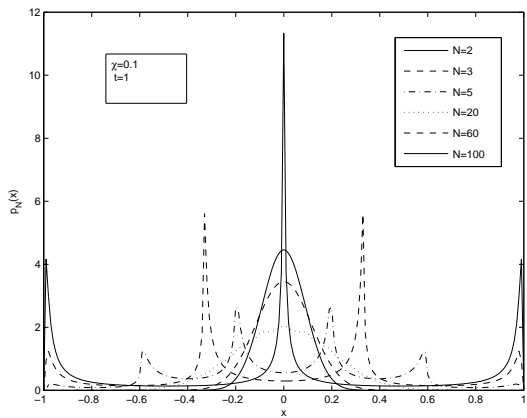}\\
  \caption{$p_N(t,x|0,0)$ against x with $\chi$=0.1 and t=1. This is the ultra-relativistic case.}\label{pxchi0_1}
\end{figure}

\begin{figure}
  % Requires \usepackage{graphicx}
  \includegraphics[width=7cm]{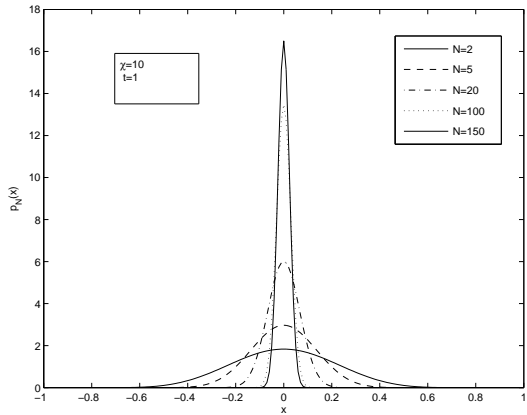}\\
  \caption{$p_N(t,x|0,0)$ against x with $\chi$=10. We have used J\"{u}ttner's function. This is the non-relativistic case. However the results are still different from (\ref{p_N_maxwell}) which is independent of N. Here large N sharpens the peak because the support of $f_J(v)$ is finite.}\label{pxchi10}
\end{figure}

\begin{figure}
  % Requires \usepackage{graphicx}
  \includegraphics[width=7cm]{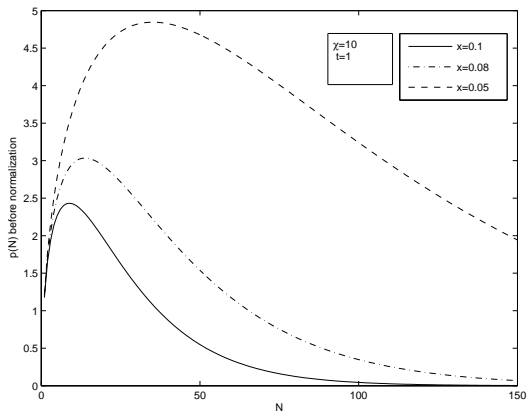}\\
  \caption{The probability distribution among different values of N before normalization. The three curves are of x=0.1, 0.08, 0.05, respectively. One can see that these curves are similar to the curves of the lognormal distributions(See Fig.\ref{lognormal}).}\label{pN}
\end{figure}

\subsection{$P_N(t,x|0,x_0)$ against x}

The formulae we use for numerical calculation have been presented in
Section \ref{RWM}. In the following, we use J\"{u}ttner's function
and we set $x_0$=0.

For $p_N(t,x|0,0)$ agaist $x$, there are three parameters: $\chi$
for the velocity distribution; $t$ for time of the diffusing
process; $N$ for the kind of the diffusing pathes.

We give our results in Fig.(\ref{pxchi0_1}) and Fig.(\ref{pxchi10}).
In Fig.(\ref{pxchi0_1}), the value of $\chi$ is 0.1. The $\delta$
peak in the middle belongs to the curve of $p_N$ with N=2. This
curve has another two peaks at $\pm$1. The two second highest peaks
which are symmetrically located belong to the curve of N=3. This
curve has low sharp peaks at $\pm$1 too. For N=4, there are 3 peaks
symmetrically located besides the peaks at $\pm$1. This case is not
shown in the picture. For large N, the $p_N(t,x|0,0)$ are
Gaussian-like, which is predicted by CLT.

In Fig.(\ref{pxchi10}), $\chi=10$. This is the case of low-velocity.
The peaks are more sharpened. The effective velocity of the
diffusing particle is small and the particle is more confined in its
neighborhood. And with N increasing the peak sharpens. This is
different with the result (\ref{p_N_maxwell}) which comes from
Maxwell's function for the PDF of velocities. J\"{u}ttner's function
with large $\chi$ applies to the low velocity situations as well as
Maxwell's function, but since the support of J\"{u}ttner's function
is finite, the causal structure is maintained.

\subsection{$P_N(t,x|0,x_0)$ against N}

For $p_N(t,x|0,0)$ against N, the parameters are $\chi$, $t$ and
$x$.

In the ultra-relativistic situation and in the region with $x$ far
away from zero, the distribution among different values of N will be
characterized by sharp peaks. Because in the region with $x$ far
away from zero, $p_N(t,x|0,0)$ is characterized by sharp peaks (See
Fig.(\ref{pxchi0_1})). If the parameter $x$ is fallen into a peak of
some $p_N(t,x|0,0)$, the probability distribution to this value of N
will be very large, and those to others will be very small.

As for the case with $x$ near zero, contributions to the total
transition PDF (the sum of $p_N(t,x|0,0)$'s with different values of
N) come mostly from large N. However, for large N, $p_N(t,x|0,0)$ is
gaussian-like. So we just consider the large $\chi$ cases in which
$p_N(t,x|0,0)$'s are Gaussian-like.

We give the results in Fig.(\ref{pN}). From Fig.(\ref{pN}), one can
see that the curves in this picture are much similar to the curves
of the lognormal distributions in Fig.(\ref{lognormal}). Lognormal
distribution has been widely used in chemistry, biology, ecology,
social sciences and economics, and many other fields \cite{LND}. We
propose that the probability distribution among different values of
N is a lognormal distribution.

\section{Conclusions}

In this letter, we have considered random walk models of diffusion
processes. One of our results is that there will be a causal Zeno
effect in causal diffusions. As for the problem of collision number,
we have found that the probabilities for the diffusing particle to
collide whatever $N-1$ time with other particles in a given time
interval will be the same if the initial position is specified but
the final position is not. For the case of that the final position
is also fixed, we have made numerical analysis and proposed that the
probability distribution among different values of N is a lognormal
distribution.

\bigskip

\noindent \bf{Acknowledgments}

This work was supported by the National Natural Science Foundation
of China(No.10275030) and Cuiying Project of Lanzhou
University(225000-582404).

\begin{appendix}

\begin{figure}
  % Requires \usepackage{graphicx}
  \includegraphics[width=7cm]{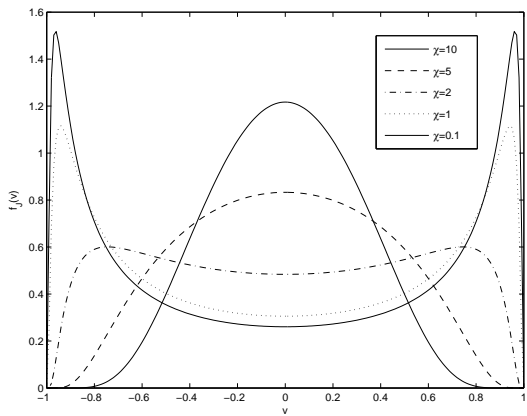}\\
  \caption{J\"{u}ttner's function. When $\chi$ is big, the temperature is small and $f_J(v)$ reduces to a Gaussian-like distribution. For small $\chi$ the peaks appear at large velocities.}\label{fJ}
\end{figure}

\begin{figure}
  % Requires \usepackage{graphicx}
  \includegraphics[width=7cm]{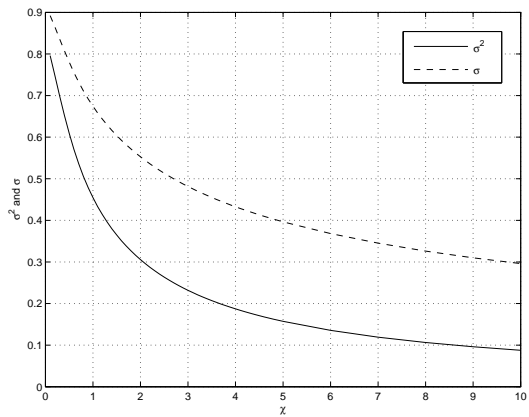}\\
  \caption{The variance and the standard deviation of J\"{u}ttner's distribution, i.e. $\sigma^2$ and $\sigma$, respectively, against the parameter $\chi$.}\label{sigma2_chi}
\end{figure}

\begin{figure}
  % Requires \usepackage{graphicx}
  \includegraphics[width=7cm]{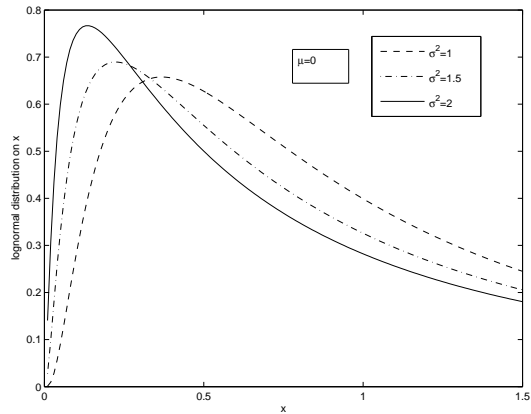}\\
  \caption{Lognormal distribution}\label{lognormal}
\end{figure}

\section{J\"{u}ttner's distribution}

J\"{u}ttner's distribution in 1-dimension space (see Fig.(\ref{fJ})
and Fig.(\ref{sigma2_chi})) is
\begin{equation}\label{f_J_1}
    f_{Jv}(v;m,\beta)=\frac{1}{2K_1(\chi)}\gamma(v)^{3}e^{-\chi\gamma(v)},
\end{equation}
where $\chi=\beta m$, $K_1(\chi)$ is a modified Bessel function of
the second kind.

The mean is 0 and the variance $\sigma^2$ is
\begin{eqnarray}
% \nonumber to remove numbering (before each equation)
  \sigma^2 =&& 1-\frac{\int dv \gamma(v) e^{-\chi\gamma(v)}}{\int dv \gamma(v)^3
  e^{-\chi\gamma(v)}} \nonumber \\
  =&& 1-\frac{1}{2K_1(\chi)}\int_{-1}^1 dv \gamma(v) e^{-\chi\gamma(v)}.
\end{eqnarray}

\section{Lognormal distribution}

Let $g_{\mu,\sigma^2}(x)$ be Gauss's normal distribution,
\begin{equation}\label{GD}
    g_{\mu,\sigma^2}(x)=\frac{1}{\sqrt{2\pi
    \sigma^2}}e^{-\frac{(x-\mu)^2}{2\sigma^2}},
\end{equation}
then
\begin{equation}
    f(x)=\left\{
           \begin{array}{ll}
             \frac{1}{x}g_{\mu,\sigma^2}(logx), & \hbox{if x $>$ 0;} \\
             0, & \hbox{if x $\leq$ 0.}
           \end{array}
         \right.
\end{equation}
is called the Lognormal distribution with parameters $\mu, \sigma^2
(-\infty<\mu<\infty,0<\sigma^2<\infty)$ \cite{CLT,LND,wikiLND} (see
Fig.(\ref{lognormal})).

\end{appendix}

%\end{CJK*}

\end{document}